\documentclass{article}
\usepackage{amsmath,amsfonts}
\usepackage{algorithmic}
\usepackage{comment}
\usepackage{graphicx}
\usepackage{textcomp}
\usepackage{xcolor}
\usepackage{subfiles}

\usepackage{xurl}
\usepackage{array}
\usepackage{multirow}
\usepackage{xspace}
\usepackage{enumitem}
\usepackage{hyperref}
\setlist[itemize]{leftmargin=4mm}
\usepackage{footmisc}
\usepackage{tabularx} 

\usepackage{fancyhdr}

\newcolumntype{L}{>{\centering\arraybackslash}m{1.5cm}}

\newcommand{\Sys}{Carbon Filter\xspace}

\AtBeginDocument{%
 \providecommand\BibTeX{{%
   \normalfont B\kern-0.5em{\scshape i\kern-0.25em b}\kern-0.8em\TeX}}}

\begin{document}

\title{\Sys: Real-time Alert Triage Using Large Scale Clustering and Fast Search}



\author{
Jonathan Oliver\textsuperscript{1}            (jonathan.oliver@broadcom.com)  \\
Raghav Batta\textsuperscript{1}                 (raghav.batta@broadcom.com)       \\
Adam Bates\textsuperscript{2}                 (batesa@illinois.edu)       \\
Muhammad Adil Inam\textsuperscript{2}                 (mainam2@illinois.edu)       \\
Shelly Mehta\textsuperscript{1}                 (shelly.mehta@broadcom.com)       \\
Shugao Xia\textsuperscript{1}                 (shugao.xia@broadcom.com)       \\
	\\
        1. Broadcom Inc., Palo Alto, CA 94304, USA. \\
        2. University of Illinois at Urbana-Champaign, IL 61820, USA. \\
}

\maketitle

\begin{abstract}

``Alert fatigue'' is one of the biggest challenges faced by the Security Operations Center (SOC) today, with analysts spending more than half of their time reviewing false alerts.
Endpoint detection products raise alerts by pattern matching on event telemetry against behavioral rules that describe potentially malicious behavior, but can suffer from high false positives that distract from actual attacks.
While alert triage techniques based on data provenance may show promise, these techniques can take over a minute to inspect a single alert, while EDR customers may face tens of millions of alerts per day; the current reality is that these approaches aren't nearly scalable enough for production environments.

We present \Sys, a statistical learning based system that dramatically reduces the number of alerts analysts need to manually review. 
Our approach is based on the observation that false alert triggers can be efficiently identified and separated from suspicious behaviors by examining the process initiation context (e.g., the command line) that launched the responsible process.
Through the use of fast-search algorithms for training and inference, our approach scales to millions of alerts per day.
Through batching queries to the model, {\it we observe a theoretical maximum throughput of 20 million alerts per hour}.
Based on the analysis of tens of million alerts from customer deployments, our
solution resulted in a 6-fold improvement in the Signal-to-Noise ratio without compromising
on alert triage performance.
\end{abstract}



\section{Introduction}\label{sec:intro}

Endpoint Detection \& Response (EDR) products continuously record 
  and store endpoint event activity, enabling security analysts to hunt 
  threats\cite{eedr}.
EDRs analyze the endpoint telemetry
  stream in real time for signs of potentially malicious activity.
EDRs are the workhorse of the endpoint layer for enterprise security,
  representing a 3 billion dollar market \cite{edr-market}
  within the security product ecosystem.

 Surprisingly, AI is not typically relied upon for this critical task;
  the endpoints of a single organization can combine to generate trillions of events (e.g., system calls) per day, making the training and inference costs of AI prohibitively costly.
Instead,  EDRs employ
a set of detection queries that
  encode likely malicious behaviors, enabling efficient pattern matching
  in a rule/heuristic-based intrusion detection model.
By annotating detection queries with techniques from the MITRE ATT\&CK framework \cite{MitreAttackFramework},
these alerts are intrinsically explainable, providing context
to analysts as they hunt.  

However, EDRs are also a major contributor to the growing ``alert fatigue'' problem --
 security products are prone to generating high volumes of alerts.
Industry reports observe that sufficiently-large organizations will face
  {\it at least} tens of thousands of alerts per day, the majority
  of which are false alerts \cite{study1, study2, fireeye}.
Among our customers,
  some organizations generate {\it tens of millions of alerts}
 of alerts per day; in an extreme case,
 a large customer with many tens of thousands of endpoints faced
 {\it 38 million to 82 million alerts each day} over a one week period.
  Analysts are still forced to identify attacks
  from a set of high volume, low severity
  making it impossible to examine all alerts \cite{agyepong2020towards,Sopan_2018}.
The reasons for high alert volumes and poor signal-to-noise ratio are numerous.
For example, attackers have become more adept at leveraging legitimate system utilities, 
  so-called "Living-off-the-land" attacks \cite{ongun2021living}, 
  blurring the boundaries between malicious and benign activity.
Further, the MITRE ATT\&CK matrix \cite{MitreAttackFramework} 
  is also populated with misuse of legitimate programs,
  with many attack techniques that are difficult or impossible to 
  distinguish from normal network activity \cite{d2020}.

\begin{figure*}[h!]
  \centering
  \includegraphics[width=0.80\linewidth]{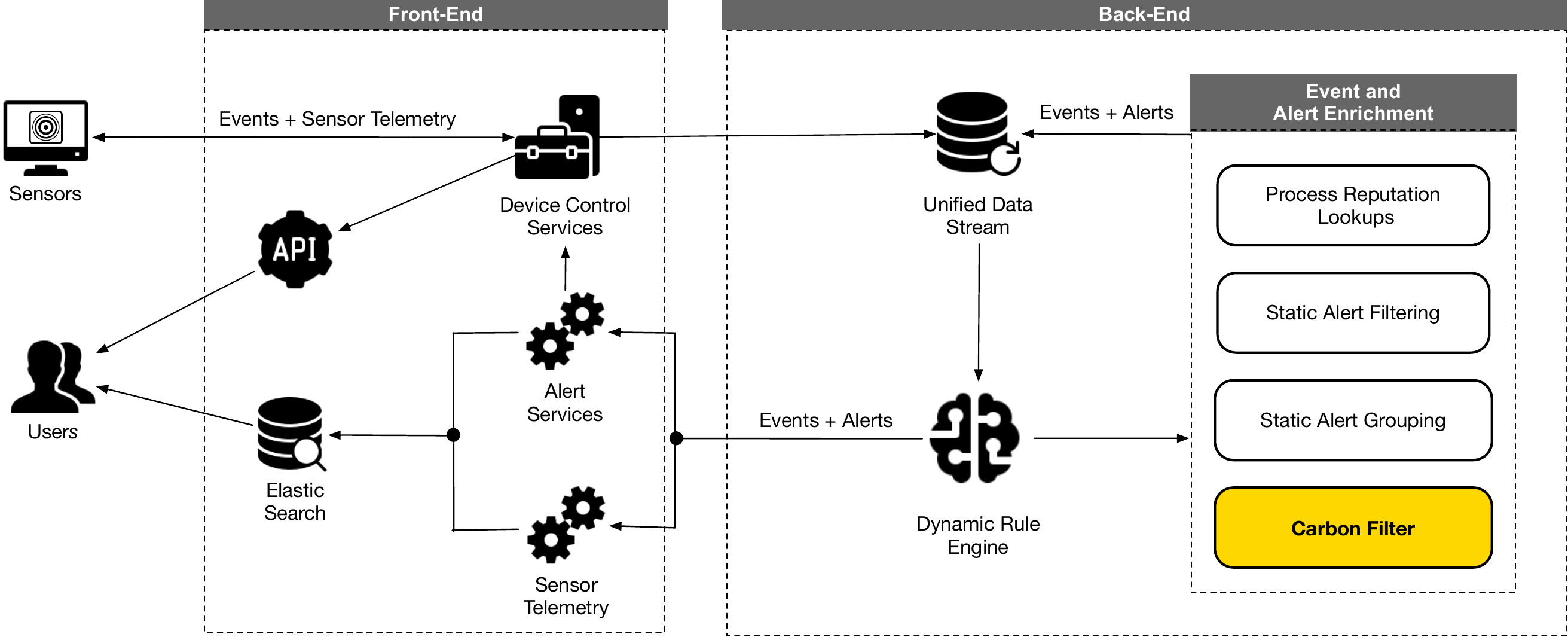}
  \caption{A simplified overview of a cloud-based EDR system based on the data collection architecture. Our approach, \Sys, enriches the alert stream by 
  classifying alerts according to their process initiation (e.g., command line) contexts.}
  \label{FigSADE}
\end{figure*}

Encouragingly, 
  recent work has begun to recognize {\it alert triage} -- 
  methods of (de)prioritizing alerts based on their perceived risk --
  as an important security task.
Most notably, data provenance analysis has been demonstrated to be 
  effective at identifying many of the false alerts in an alert stream \cite{NODOZE, hbm2020, mge+2019}.
These approaches examine the historical causal relationships of each process
  that causes an alert to fire, effectively examining the suspiciousness of past activity
  to determine the suspiciousness of the current execution context.
On small evaluation testbeds (tens to hundreds of machines), 
  provenance-based triage has been shown to be capable of filtering out 84\% \cite{NODOZE} to 97\% \cite{hbm2020} of false alerts, 
  a tremendous improvement in signal-to-noise ratio.
Unfortunately,
  {\it we concluded that provenance-based triage is not nearly scalable
  enough to support the needs of a commercial EDR system in its present form.}
Because provenance needs to retrieve and process many historic events
  for triage, it typically takes tens of seconds (if not minutes) to 
  analyze a single alert \cite{NODOZE, hbm2020}.
What is needed is a technique that approaches the efficacy of provenance-based triage
  with a throughput on the order of thousands of alerts per second.

In this work, we propose that alerts can actually be triaged
  by examining processes' {\it immediate} (not historic) execution context in the form of the causal initiator of the process. Typically this initiator will be a command line.\footnote{There are multiple possible initiators for a process. The most common process initiator is the command line which started the process. Other common process initiators include a user clicking an icon on the desktop, process creation from within another process and scheduled tasks. We construct a `proxy' command line for these initiators.\label{foot:initiator}}
Even when examining alerts fired by the same program,
  command line arguments can be used to delineate 
  false alert sources from truly suspicious activity.
In fact, we predict that a large proportion of global 
  false alerts are the result of just a handful of
  process command line contexts.
Our solution, \Sys, performs similarity-preserving hashing of every 
  process command line that triggers an alert, then identifies
  large clusters in which nearly-identical command lines have
  caused an inordinate number of alerts.
We apply approximate nearest neighbor searches
  to classify future alerts.
For alerts which fall into a cluster, we apply outlier detection to detect suspicious activity.
Finally, for alerts that do not fall into a cluster,
  we introduce an efficient alert scoring model based on 
  examining the frequency of a command line's occurrence 
  on the device, organization, and global levels.
By prioritizing efficient analysis and avoiding costly provenance queries,
  we are able to triage $\sim 5.5K$ alerts per second on a single multi-core machine.

We evaluate \Sys using over
  288 million alerts generated by our customers,
  including over 10,000 labeled alerts that were verified
  by experts in the Threat Analysis Unit (TAU)
  and Managed Detection and Response (MDR) groups.
We first demonstrate that 102 million alerts map to $763K$ unique similarity digests, which reduce to a compact representation of $16K$ clusters.
A relatively small number of clusters provide an explanation for just over half of the alerts (54.5\% of the 102 million alerts used in training belonged to just $1,638$ clusters).
Finally, we demonstrate that \Sys is able to filter out 82\% of false alerts 
  while mis-classifying 0.1\% or less of malicious alerts, 
  resulting in a 6-fold improvement in the signal-to-noise
  ratio of the alert stream.
{\it \Sys is already deployed in a
  production environment and is already helping customers
  hunt threats.}

The key contributions of this paper are as follows:

\begin{itemize}
	\item {\it Triage millions of alerts per hour.} We demonstrate a scalable method of alert triage
    through granular clustering of command lines, prioritizing analysis throughput over noise reduction.

    \item {\it Global-scale Evaluation.} We demonstrate the efficacy of our approach on a global snapshot of our 
    customer data, comprised of hundreds of millions of alerts.

    \item {\it Open Evaluation to Prior Work.}
    While we are unable to release our proprietary data,
    we replicate our results on the open-source ATLASv2 dataset.
    Here, we compare our approach to the NoDoze \cite{NODOZE}
    and RapSheet \cite{hbm2020} provenance-based triage techniques.
    To facilitate comparison, we will release
    our scores for the ATLASv2 alerts upon publication.

\end{itemize}

The rest of this paper is organized as follows.
In Section \ref{sec:background}, 
  we provide an overview of Endpoint Detection \& Response
  architectures and highlight limitations of current
  alert management strategies.
Section \ref{sec:design} presents the design and
  implementation of \Sys.
In Section \ref{sec:evaluation},
  we report on the performance of \Sys 
  and compare our approach to provenance-based methods.
We provide additional commentary on related work in 
  Section \ref{sec:related_work},
  discuss future directions in Section \ref{sec:future_work},
  and conclude in Section \ref{sec:conclusion_future_work}.


\section{Background \& Motivation}
\label{sec:background}

\subsection{Endpoint Detection and Response}

\label{sec:cb_data}

Endpoint Detection and Response (EDR) products
continuously monitor and record the activities and events taking place on end-user devices to detect and promptly respond to cyber threats. In this section, we provide a high-level overview of the infrastructure, a system that embodies industry-standard practices in the domain. The architecture shown in Figure~\ref{FigSADE}, is structured such that the front-end interface is accessible to customers through APIs, while the complexities of the backend remain invisible to end-users. We go over the individual components of the architecture in detail below.

Sensors are deployed on target machines, where they scan the system to generate a continuous stream of events. Additionally, the sensors monitor for the introduction of new executable content. The event streams generated by the  sensors are passed to the device control service, where specific device-based configurations are applied to the event stream. Subsequently, the updated event stream, augmented with device-specific metadata, is transmitted to the backend infrastructure.

The event stream proceeds to the Unified Data Stream (UDS), which serves as the data processing platform at the backend. From the UDS, the event stream is sent over to the Dynamic Rule Engine (DRE). The DRE, referred to as the "backend brain", is responsible for applying expert-defined behavioral rules to the event telemetry. These behavioral rules are founded on the concept of "state changes" as described in the MITRE Attack Framework~\cite{MitreAttackFramework}. This approach enhances the resilience of the system against evasion tactics compared to rules based on procedural descriptions of actions. For instance, technique T1037.004~\cite{MitreEtcRc} establishes persistence by modifying RC scripts referenced during UNIX startup. The behavioral EDR rule is defined as follows:

\begin{verbatim}
IF File-modified(/etc/rc.d, /etc/init.d, /etc/rc.local)
    AND Is-Platform(Linux, macOS)
THEN FIRE
\end{verbatim}

Irrespective of the method employed to acquire the necessary privileges for this state change, any events involving alterations to these files trigger alerts. The DRE also incorporates various event and alert enrichment techniques into the event telemetry. These techniques include process reputation lookups, static alert filtering, and static alert grouping. The details of each technique along with their limitations are discussed in Section~\ref{sec:limitations}.

Following the application of behavioral rules and alert enrichment techniques, the resultant security alerts are stored within the Carbon Black Cloud (CBC) and transmitted to the alert and sensor telemetry services at the front-end. If a known malicious process is identified on the endpoint, alert services notify the endpoint sensor to terminate the process. Moreover, analysts operating at the front-end of the infrastructure can utilize Elastic Search to query and retrieve specific alert or telemetry data for further analysis.

\subsection{Limitations of Prior Approaches}
\label{sec:limitations}

In this section, we discuss various existing alert triage techniques along with their inherent limitations. As discussed earlier, the DRE employs a set of event and alert enrichment techniques to both enhance alert metadata and mitigate false alerts. Figure~\ref{FigSADE} provides an overview of some widely practiced strategies within the cybersecurity industry, including process reputation lookups, static rule-based filtering, and static alert grouping. Additionally, we discuss data provenance and causality-based approaches for alert triage~\cite{NODOZE, hbm2020, mge+2019}.

\par \textbf{Process Reputation Lookups} - 
EDR products assess the reputation of a running process, considering factors such as the reputation of the process's primary executable and its parent process.

Typically, process reputation lookups rely on a repository of known processes and malware signatures. This methodology identifies known attack signatures, however, it struggles with more sophisticated threats, including living-off-the-land techniques.

\par \textbf{Static Alert Filtering} - 
Due to high alert volume, EDR products incorporate predefined rules to filter out false alerts. EDR products may modify and adjust existing behavioral rules or introduce new filtering rules based on the ongoing alert generation patterns.

While static alert filtering can reduce the volume of alerts, it introduces the risk of malicious alerts being filtered out. There is a possibility that attackers may employ the same behavioral tactics targeted by the filtering and not raise any alerts. 

\par \textbf{Static Alert Grouping} - 
EDR products typically offer the capability of static alert grouping, enabling the correlation of alerts by various event attributes, such as originating hosts, alert types, or severity levels. While these techniques facilitate the organization and categorization of alerts, they do not directly address the core challenge of reducing false positives.

\par \textbf{Provenance-based Approaches} - 
Provenance-based strategies delve into the historical causal relationship of processes leading to an alert by analyzing the suspiciousness of prior activities~\cite{NODOZE, hbm2020, mge+2019, kairos}, a process that is computationally intensive due to the need to process extensive historical event data. While these approaches have shown promise in small evaluation testbeds, their scalability in commercial EDR systems remains a significant challenge.

One of the critical limitations of provenance-based approaches is their reliance on the complete event stream for provenance-graph generation. In our production environment, the ratio of events to alerts stands at 1500:1, making it inherently difficult to scale provenance analysis due to the complexity of managing such event streams. Although Kairos represents a notable advancement in processing speed while utilizing provenance analysis~\cite{kairos}, it is only capable of handling 8 alerts per second\footnote{
Kairos achieving an alert throughput of 8 alerts per second is derived from applying the 1500:1 event-to-alert ratio to its documented performance of 11k events per second. }. In contrast, as shown in Table \ref{tab:ann-quantification}, our detection engine processes approximately 1,317 alerts per second based on actual customer data (with a max. load throughput of 5,566 alerts per second), highlighting the substantial gap in scalability.

\section{\Sys Methodology}
\label{sec:design}


\subsection{Design Goals}
\label{subsec:design-goals}
We set out to design an alert classification system that satisfies
  the following requirements.
\begin{enumerate}[leftmargin=*]
\item {\bf Highly Scalable.} The sensors running on customer endpoints can generate tens of millions of alerts per day, a volume that far exceeds the capacity of 
prior triage approaches in the literature. Our solution must be able to support massive streams of alerts during both training phase and inference (triage) phase.

\item {\bf Effective with Limited Labeled Data.}
Scarcity of labeled data is a major challenge for the cybersecurity industry, and presents a unique challenge for applying ML to cybersecurity \cite{CIOArticle}.
While it is possible to have SOC analysts manually triage alerts for labeled validation data, this is a labor intensive task that does not scale.
Our solution must therefore effectively filter out false alerts from the alert stream with access to only limited labeled samples.

\item {\bf Resilient to Class Imbalance}
The majority of the activity that happens on an endpoint is legitimate and benign, with attacks occurring very rarely.
This creates the conditions for high false alert rates that lead to alert fatigue, a problem that spans across the majority of organizations in the security industry \cite{ESGReport}. In addition, this also leads to the potential for bias in ML models
trained on alert data. Our solution must be able to effectively differentiate false alerts from malicious alerts in spite of this class imbalance.


\end{enumerate}


\subsection{Application Architecture}

\begin{figure}[t!]
    \centering
    \includegraphics[width=\linewidth]{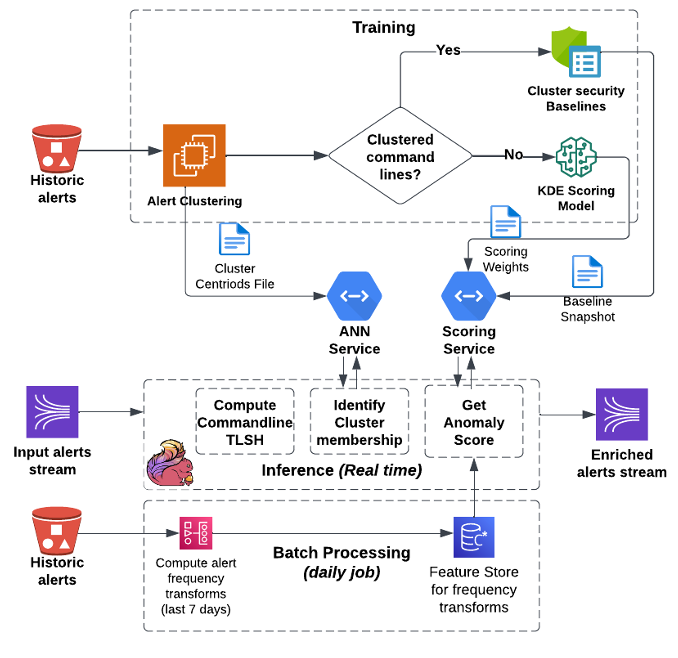}
    \caption{\Sys's training and inference phase.}
    \label{Overview}
\end{figure}

\Sys interposes on the Unified Data Stream (UDS) explained in Figure \ref{FigSADE} as an alternative enrichment mechanism. The overall workflow of \Sys is shown in Figure~\ref{Overview}. 

\par \textbf{Training} - During a training phase, historic alerts are collected from the S3 \cite{s3} data storage. The command line for the process that triggered the alert are pre-processed and vectorized using a computationally-efficient similarity-preserving hash (TLSH). A clustering algorithm then identifies clusters, and we collect baseline statistics of 28 security features within each cluster. For the unclustered alerts, a KDE based scoring model is trained using the process command line and parent-child process paths.

\par \textbf{Inference} - During the inference phase, \Sys leverages Apache Flink \cite{Flink} as a scalable streaming framework. The real-time alerts are transferred to the Flink pipeline from UDS via AWS Kinesis \cite{Kinesis} streams in the form of Protobuf objects \cite{Protobuf}. The alerts are pre-processed, TLSH is computed for their command lines and the ANN endpoint is queried to identify the cluster ID for the alert. Once the cluster ID is determined, the enriched alert is sent via a post request to the scoring endpoint which returns the score for both the clustered and unclustered alerts. The enriched protobuf objects are then transferred back to the UDS Kinesis stream. 

\par \textbf{Batch Processing} - The batch processing (daily) job for computing the alert frequencies over the last 7 days (refer $\S$ \ref{subsec:anomaly_score}) utilizes Apache Airflow \cite{Airflow}. The frequency transforms are saved into Cassandra \cite{Cassandra} data storage library and is queried by the Flink pipeline for unclustered alerts before sending the request to scoring service.

The remainder of this section will explain the major components of \Sys. We first explain how alert processes' command lines are vectorized ($\S$\ref{SecUsingTLSHforClustering}) and clustered ($\S$\ref{HACTClustering}) in the training phase.
We then outline the mechanisms of the inference (test) phase, specifically \Sys' use of Approximate Nearest Neighbor (ANN) cluster search ($\S$\ref{sec:ANN}) and the alert scoring model used to triage alerts that fall outside of a cluster ($\S$\ref{subsec:anomaly_score}).

\label{sec:system_overview}


\subsection{Process Command Line Hashing}
\label{SecUsingTLSHforClustering}

\Sys represents alerts using the command line arguments of the process that caused the alert to trigger.\footref{foot:initiator}
To understand the intuition behind this approach, let us consider a simple program like Unix utility for which all processes with the same command line would raise the same alert. 

However, even for more complex and feature-rich programs, the command line provides meaningful information as to how the program is being use. In particular, in order for ``Living off the Land'' attacks to be successful, we expect that the attacker must manipulate the behavior of the legitimate program by deviating from the standard command line arguments that appear during typical use. Therefore, the command line arguments used to launch a program should effectively differentiate malicious alerts from the false alerts associated with typical use.

Our system vectorizes process command line arguments using similarity digests \cite{oliver2020hac}.
Popular similarity digests include Ssdeep \cite{kornblum2006identifying}, Sdhash \cite{roussev2010data}, LZJD \cite{lzj}, and TLSH\cite{oliver2013TLSH}.
SSdeep and Sdhash do not support input lengths we would commonly expect from command lines, with minimum input lengths of 4096 bytes and 512 bytes respectively.
Under the extreme scales of our execution environments, some digests are not fast enough for our purposes (see \cite{c28, DellAmico2019}).
Given its support for smaller inputs, fast performance, effectiveness in related cybersecurity tasks and resistance to evasion \cite{oliver2020hac} we choose TLSH for command lines. Refer to the supplmental material ($\S$\ref{subsec:tlsh_dist_example}) for an illustrative example of TLSH distance calculations on command lines.


A small proportion (roughly 2\% of alerts) have command lines shorter than 50 bytes. For these alerts we apply some prepossessing steps to create transformed command lines with at least 50 bytes.

\subsection{Alert Clustering}
\label{HACTClustering}

The goal of clustering alerts, represented by their process command lines,\footref{foot:initiator} is to identify granular groups of alerts with extremely low entropy for multiple security features. Thus if the groups are sufficiently large in size, the various features within the group can be used to form a baseline of expected behaviour. This is consistent with standard operations within the SOC.

While a variety of clustering algorithms would be effective for this task, we choose the Hierarchical Agglomerative Clustering (HAC-T) algorithm \cite{oliver2020hac} using the TLSH distance function.
The ouput of HAC-T clustering is a list of cluster centroids in the form of a TLSH, the radius of the cluster
and number of points that belong to each cluster.
HAC-T stops growing a cluster when the radius of the cluster is greater than our threshold distance, $CDist$.
We set this threshold through iterative experimentation over large volumes of alert data, with the goal being to identify a threshold that creates a large number of extremely low-entropy clusters as described in the supplemental material ($\S$ \ref{subsec:ChoosingoptimalCDist}).

We note that clustering the alerts has the additional benefit of making the overall system more scalable during the inference phase. During inference, we need to determine which cluster a new alert belongs to, limiting the search to the cluster centroids for the closest match, instead of the entire training set. Typically the number of cluster centroids will be of the logarithmic order of the number of data points in the training set.

\subsection{Approximate Nearest Neighbor Search}
\label{sec:ANN}


The customers average $921$ $\pm$ $396$ (standard deviation) alerts per second 
that need to be associated to one of the clusters for the product area in focus.
Therefore a scalable approach is needed to 
  classify new process command lines 
  in near real time.
We employ an approximate nearest neighbor based search 
  to identify 
  the nearest cluster centroid to which the newly generated alert belongs.
This is both scalable and a flexible architecture \footnote{
We could in future change the TLSH vector to another embedding vectors (computed
through deep learning or traditional machine learning techniques) while leaving the overall
architecture intact.}.

Typical approximate nearest neighbor (ANN) methods fall in one of the following
categories - Tree based data structures like ANNOY~\cite{annoy},
 Neighborhood graphs like HNSW~\cite{malkov2018efficient} and NN-Descent~\cite{DongCL11}.
NN-Descent for example constructs a nearest neighbor graph and then
performs search for approximate nearest neighbors. This graph is built by connecting each
data point to k-most similar points from the cluster centroid dataset. It act as an index
structure over which we perform breadth-first search and only keeping the best k-points
at any points in the traversal.




We used NN-Descent algorithm with custom TLSH distance for finding the appropriate cluster. We experimented with TLSH distance function \cite{TLSHSoftware} in python both as a Numba~\cite{lam2015numba} function and a Cython~\cite{behnel2011cython} function. The Numba compiled distance function achieves almost twice the speedup compares to the Cython one.


\subsection{Anomaly Detection for Clustered Alerts}

It is not justified to classify an alert as being a false alert purely based on its command line.
Attackers will use a variety of tactics to mimic legitimate activity such as performing malicious activities with legitimate software ("Living-off-the-land"),
or renaming malicious programs to the names of legitimate software. Therefore, we apply outlier detection to all alerts which fall within a cluster.

The outlier detection works in the following way.
We have clustered the training set which resulted in high homogeneity of the security features associated with alerts (refer $\S$\ref{HACTClustering}).
During training, each cluster has a profile constructed for each of the 28 security features including MITRE TTPs, process reputation, privilege escalation, digital signature states, process path, and process user names, etc. 
During the inference phase, we apply outlier analysis\footnote{The outlier analysis is simply the process of identifying values that occur less frequently than a specified proportion of the population (such as 1 in 1000).}
to the security features.
Thus an alert will only by classified as a false alert when all the security features are consistent with the baseline defined by the alert's associated cluster.

We fully anticipate that, over time, common attack techniques will occur often enough to form their own clusters within our model.
To account for this, we continue to monitor the correctness of the clusters through review of feedback by customers and our expert analysts. If a malicious alert is found to be present in a cluster, it is immediately marked as contaminated at which point all future alerts that fall within the cluster proceed to the alert scoring phase described below.
This makes correction of each malicious cluster a one-time effort as it forms.

\subsection{Anomaly Detection for Unclustered Alerts}
\label{subsec:anomaly_score}

If an alert does not belong to a cluster, 
  \Sys falls back to a more traditional
  anomaly-based scoring system to prioritize
  the alert for analysts.
Our approach is to consider the frequency of a given
  alert's occurrence with regards to two aspects of the alert:
  (1) its process command line,
  as well as (2) parent-child process path\footnote{
The parent-child process path features proved useful to identify anomalies related to living off the land activity.}.

To calculate the anomaly score,
 we use a Kernel Density Estimate (KDE) statistical model \cite{KDE2020}.
We calculate alert frequencies for a 7 day window for six features of the alert,
  the process command line\footnote{Due to the variations in command lines, we use the clustering model to calculate these frequencies.} (on the device, within the organization and globally) and the parent-child process path (on the device, within the organization and globally).
This gives us
$f_1,~ f_2,~ f_3,~ f_4,~ f_5,~ f_6$, where $f_1$ and $f_2$ are device based,  $f_3$ and $f_4$ are organizationally based and $f_5$ and $f_6$ are the global frequencies.
We apply a logarithmic transformation to each frequency:
\begin{equation}
F_i = \frac{1}{1 + \log{(1 + f_i)}}
\end{equation}
We calculate a scaling factor for organizational and global features:
\begin{equation}
S_1 = S_2 = 1; ~~~~~
S_3 = S_4 = \log{\frac{n_{od}}{n_{o}}}; ~~~~~
S_5 = S_6 = \log{\frac{n_{go}}{n_{g}}}
\end{equation}
where $n_{o}$ is the number of the devices where the alert occurs, $n_{od}$ is the total number of devices in the organization, $n_{g}$ is the number of the organizations where the alert occurs  and $n_{go}$ is the total number of organizations.  The scaling factor is added to organizational and global features to give higher weight when an alert occurs across more devices as compared to the same frequency of alerts occurring on fewer devices.
The overall anomaly score is then
\begin{equation}
Score ~=~ \sum_{i=1}^{6}{ w_i \times F_i \times S_i}
\end{equation}
where the weights $w_i$ are estimated by maximizing the ROC AUC using the data described in Section \ref{subsec:evaluation_label}.


\section{Evaluation}\label{sec:evaluation}



Unless otherwise noted, 
  all experiments were conducted on an AWS hosted EC2 machine with the following configurations:
CPU Model name: Intel(R) Xeon(R) Platinum 8175M CPU @ 2.50GHz, L1d cache: 32K, L1i cache: 32K, L2 cache: 1024K, L3 cache: 33792K.

\subsection{Experimental Data}\label{subsec:evaluation_label}

Evaluating the performance of \Sys requires us to collect
  large amount of alert data with ground truth labels
  for both {\it malicious and false alerts}.
  As such labeled data is often not available,
we enlisted the help of two
  expert analyst groups.

\textbf{TAU Dataset:}
The Threat Analysis Unit (TAU) team authors and maintains the rules running on the telemetry data captured by the sensors on customer endpoints.
Over a 3 month period, we collected
  $7,193$ labeled alerts out of which $74$ were annotated as malicious. 
To minimize human error, we had each alert reviewed
  by multiple TAU analysts, computed the inter rater reliability score using Cohen's kappa coefficient~\cite{kappa}, and removed the alerts which had conflicting annotations.


\textbf{MDR Dataset:} The Managed Detection and Response (MDR) team actively triages alerts for customers through the MDR Console and working to prevent security attacks.Over a one month period, we collected $2,822$ labeled alerts, of which $88$ were annotated as malicious by the MDR analysts.

The above two datasets provide us with an invaluable
  snapshot of global security alert activity over the course
  of several months.
As this data is sensitive,
  it cannot be shared with the security community.
To promote open science and facilitate comparison to prior work,
  we also make use of the public ATLASv2 dataset.

\textbf{ATLASv2 Dataset:}
The {\it ATLAS Attack Engagements, Version 2} (ATLASv2) dataset~\cite{riddleatlas} represents an enriched version of the ATLAS engagement initially conducted by Alsaheel et al~\cite{alsaheel2021atlas}. The ATLASv2 dataset features an enhanced methodology that includes an extended four-day period of naturalistic benign background activity, a one-day attack period in which benign activities continue to occur, 
and the integration of additional telemetry sources, notably an EDR sensor. In total, this dataset consists of 491 alerts, among which 50 are categorized as malicious alerts.
We generated \Sys cluster labels for the ATLASv2 dataset
  and will release them upon publication.
However, 
  it should be noted that we cannot apply our scoring model 
  to this dataset because the necessary profiling information does not exist.

\subsection{Clustering Performance}

We trained the clustering model of \Sys using an additional 
  months worth of unlabeled customer data 
  comprised of almost $\sim 102$ million alerts,
  which were initiated by $\sim 33$ million unique command lines,
After applying TLSH digests and removing duplicates,
  this reduced to $\sim 763$K unique TLSHs.
The next step was to determine the clustering parameters.
The HAC-T algorithm has a threshold dist ($CDist$ in Algorithm 4 of \cite{oliver2020hac}) which plays the same role
as $\epsilon$ (or $eps$) in DBSCAN \cite{schubert2017dbscan}.


The optimal value for $CDist$ is a trade-off between the number of clusters and the homogeneity within each cluster. 
In Appendix ~\ref{subsec:ChoosingoptimalCDist}, we describe how we selected $CDist$ using an entropy criterion.

\subsubsection{Clustering Results}

Having selected $CDist ~=~ 50$,
  we clustered the $\sim 763$K unique TLSH
  digests in the training sample,
  which resulted in 16,299 clusters.
These clusters combine to account for   
  $\sim $101 million (99.1\%) of the alerts in our dataset,
  leaving only 499,946 (0.9\%) left unclustered.
The net effect of using TLSH and clustering resulted in describing 102 million alerts by 16K clusters.
This reduction is essential to the scalability
  of our end-to-end system as it allows us to more efficiently
  classify new alerts.


\subsection{ANN Search Results}\label{subsec:evaluation_ann}

\Sys's scalability crucially depends on classifying alerts in real-time, highlighting the importance of our ANN search algorithm's performance. To evaluate this, we conducted experiments varying query sizes, CPU cores, and k-neighbors, using a dataset of $\sim983$K unique TLSHs from $\sim198$ million monthly customer alerts. These TLSHs were matched against $\sim16K$ identified clusters. We evaluate the performance by computing recall, measuring the fraction of relevant retrieved results against all relevant database entries, is calculated as:
\begin{equation}
Recall = \frac{\text{Number of relevant results found}}{\text{Total relevant results in the database}}
\end{equation}

\begin{table*}[htb]
    \caption{Search times of Linear Search vs. ANN in seconds. Query batch size $=500$. Varying index sizes and K neighbor values.}
    \label{tab:ann}
    \centering
    \small
    \begin{tabular}{|c|c|c|c|c|c|c|}
	\hline
		&
    		\multicolumn{2}{|c|}{Linear Search} &  
    		\multicolumn{2}{|c|}{Single Core ANN} &  
    		\multicolumn{2}{|c|}{Multi Core ANN} \\
	\hline
      K & Index=16K & Index=100K
      & Index=16K & Index=100K
      & Index=16K & Index=100K \\
    \hline
    1 & 1.01s & 10.42s & 0.67s & 3.37s & {\bf 0.09s} & \textbf{0.38s}\\
	\hline
    3 & 1.01s & 10.42s & 1.10s & 3.79s & {\bf 0.10s} & \textbf{0.44s}\\
	\hline
    5 & 1.01s & 10.42s & 1.00s & 5.00s & {\bf 0.10s} & \textbf{0.56s}\\
	\hline
    \end{tabular}
\end{table*}

\begin{table*}[htb]
    \caption{Recall Rates for ANN. Query batch size $=500$. Varying index sizes and K neighbor values.}
    \label{tab:ann:recall}
    \centering
    \small
    \begin{tabular}{|c|c|c|}
	\hline
	K & Index=16K & Index=100K \\
	\hline
    1 & 0.96 & 0.96 \\
    3 & 0.97 & 0.96 \\
    5 & 0.97 & 0.96 \\
	\hline
    \end{tabular}
\end{table*}

\begin{table*}[htb]
    \caption{Quantification of ANN throughput to handle peak workload. Query batch size $=500$. Varying index sizes.}    
    \label{tab:ann-quantification}
    \centering
    \small
    \begin{tabular}{|c|c|c|c|c|c|}
	\hline
    \multicolumn{2}{|c|}{Total Time (ms)} & 
    \multicolumn{2}{|c|}{Time per alert (ms)} & 
    \multicolumn{2}{|c|}{Max Load handled by} \\
    \multicolumn{2}{|c|}{} & 
    \multicolumn{2}{|c|}{} & 
    \multicolumn{2}{|c|}{ANN (alerts per sec)} \\
	\hline
        Index=16K & Index=100K
      & Index=16K & Index=100K
      & Index=16K & Index=100K\\ 
    \hline
     90 & 380 &		0.18 & 0.76 &		5,566 & 1,317 \\
    \hline
    \end{tabular}
\end{table*}

The performance results are reported in Table~\ref{tab:ann} and the recall rates are reported in Table~\ref{tab:ann:recall}.
Performance tests on single and multi-core (16 cores) CPUs, with batch queries of 500 alerts, showed multi-core searches were up to $10\times$ faster than single-core and $100\times$ faster than linear searches, with minimal recall differences between index sizes of $16K$ and $100K$. Search times increased approximately fourfold when scaling the index from $16K$ to $100K$.

Quantification of the ANN search's upper limit (see Table~\ref{tab:ann-quantification}) revealed that at $16K$ index size, it could process $\sim5,566$ alerts per second, capable of handling over four times our current peak load. This indicates robust scalability, even for a $100K$ index size, supporting our real-time classification goal.

\subsection{Scoring Results}
In this section, we discuss the results of \Sys's scoring model. We also provide an
explanation of the bias-correction approach we used and our evaluation criteria.

\par \textbf{Bias-Correction:} The feedback on alerts from the TAU team and MDR analysts, predominantly skewed towards the malicious class due to a selective sampling strategy and does not accurately reflect the broader alert population. This discrepancy introduces a bias in the training dataset, potentially skewing model performance metrics and compromising the accuracy of predictions in a live environment. To mitigate this, we implemented a bias-correction technique, adjusting our dataset to more closely represent the actual distribution of alert classes within the general population. This process involved augmenting the dataset by duplicating specific alerts in proportion to their real-world occurrence rates, as illustrated in Table~\ref{feedback-tao-mdr-data}. Notably, this correction was applied solely to the final month's worth of TAU feedback, designated for testing, while the initial two months served to train the scoring model. Following bias-correction, the representation of malicious alerts significantly decreased to 0.15\% and 1.2\% for TAU and MDR data, respectively, from the original labels of 1.4\% and 3.1\%.

\begin{table}[htb]
    \caption{Distribution of test data for false and malicious classes 
    before and after bias-correction.}
    \label{feedback-tao-mdr-data}
    \centering
    
    \begin{tabular}{|c|c|c||c|c|}
    \cline{2-5}
    \multicolumn{1}{c}{} &  
    \multicolumn{2}{|c||}{Before Bias-Correction} & 
    \multicolumn{2}{c|}{After Bias-Correction} \\    
    \cline{2-5}    
    \multicolumn{1}{r|}{}
      & Malicious & False
      & Malicious & False \\ 
    \hline
    \hline
    
    TAU & 32 & 2,218 & 108 & 71,156\\
    \cline{1-5}
    MDR & 88 & 2,734 & 1,003 & 81,294\\
    \cline{1-5}
    \end{tabular}
\end{table}

\par \textbf{Evaluation Criteria:} Our primary focus is on improving the Signal-to-Noise ratio, driven by two key considerations. Firstly, the security domain prioritizes minimizing False Negatives (incorrectly dismissing a genuine threat) over reducing False Positives (erroneously flagging benign activities), emphasizing the critical nature of maintaining high Recall without sacrificing it for improved Precision. Secondly, given the pronounced imbalance between malicious and benign classes, with Precision being particularly susceptible to such disparities, our efforts concentrate on bolstering the Signal-to-Noise ratio rather than optimizing Precision alone.

\textbf{Performance on TAU and MDR Dataset:} Employing the initial 67\% of TAU feedback for model training, with the remainder and the complete MDR feedback dataset serving as the testbed, we assessed our model's efficacy. Alerts scored at or above a 0.5 anomaly threshold were classified as malicious. Our results are reported using the bias-correction (Table~\ref{feedback-tao-mdr-data}), which is a much better estimator of the performance in production scenario. The results for the TAU and the MDR datasets are shown in Table~\ref{scoring-tao-mdr-eval}. These results demonstrate that our model is able to remove $82.5\%$ of false alerts from the triage queue for TAU dataset while retaining a $100\%$ of actual malicious alerts. For the MDR dataset, our approach removes $84\%$ of false alerts while retaining $94.3\%$ of actual malicious alerts. This translates to approximately a 6-fold improvement in the Signal-to-Noise ratio.

\begin{table}[htb]
    \caption{\Sys Performance on labeled TAU and MDR data after bias-correction.}
    \label{scoring-tao-mdr-eval}
    \centering
    
    \begin{tabular}{|c|c|c||c|c|}
    \cline{2-5}
    \multicolumn{1}{c}{} &  
    \multicolumn{2}{|c||}{Retained for triage} & 
    \multicolumn{2}{c|}{Removed from triage} \\    
    \cline{2-5}    
    \multicolumn{1}{r|}{}
      & Malicious & False
      & Malicious & False \\ 
    \hline
    \hline
    
    TAU & \small{108 (100\%)} & \small{12,477 (17.5\%)} & \small{0 (0\%)} & \small{58,679 (82.5\%)}\\
    \cline{1-5}
    MDR & \small{945 (94.3\%)} & \small{12,971 (16\%)} & \small{58 (5.7\%)}  & \small{68,323 (84\%)}\\
    \cline{1-5}
    \end{tabular}
\end{table}

\subsection{Comparison to Provenance Baselines}
\label{sec:compare}

To address scalability concerns with provenance-based approaches in commercial EDR systems, we conduct a comparative analysis of \Sys against two notable methods: NoDoze~\cite{NODOZE} and RapSheet~\cite{hbm2020}. NoDoze analyzes event frequency and data flows, while RapSheet focuses on alert metadata and temporal sequencing within the MITRE kill chain context.

\par \textbf{Time Overhead Comparison:}
A direct comparison between the offline nature of NoDoze and RapSheet with \Sys presented challenges. For fairness, we adapted both methods for streaming and on-demand functionality. Moreover, we utilize linear search instead of multi-core ANN in our approach for a fair comparison. The computational overhead, defined by the time required to process a single alert, reveals that \Sys significantly outperforms NoDoze and RapSheet. Specifically, \Sys is 5,064 times faster than NoDoze and 26,723 times faster than RapSheet, as detailed in Table~\ref{tab:computational-overhead}. This stark difference highlights the limitations of traditional provenance-based methods in scalability and underscores the efficiency of \Sys.

\begin{table}[htb]
\centering
\small 
\caption{Computational overhead comparison of \Sys with two provenance-based alert-triage approaches.}
\label{tab:computational-overhead}
\begin{tabularx}{\columnwidth}{|X|c|c|}
\hline
\textbf{System} & \textbf{Time per Alert (s)} & \textbf{Total Time (s)} \\
\hline
Carbon Filter (Linear) & 0.002 & 0.99\\ \hline
NoDoze & 10.23 (5,064X) & 5024.2 (5,064X) \\ \hline
RapSheet & 53.98 (26,723X) & 26503.6 (26,723X) \\ \hline
Carbon + NoDoze & 10.23 (5,064X) & 1780.02 (1,798X) \\ \hline
Carbon + RapSheet & 53.98 (26,723X) & 9392.52 (9,486X) \\ \hline
\end{tabularx}
\end{table}

\par
\begin{figure}[htb]
    \centering
    \includegraphics[width=0.9\linewidth]{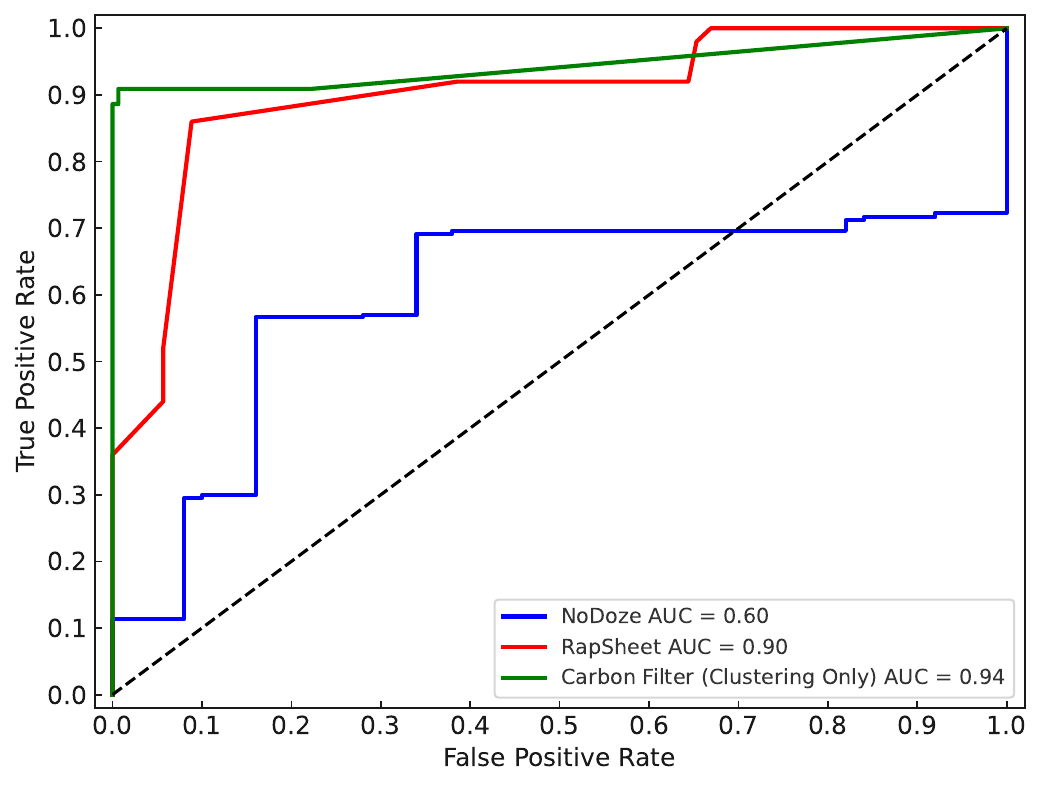}
    \caption{AuC comparison of \Sys with provenance-based approaches on the ATLASv2 dataset.}
    \label{figauc}
\end{figure}

 \textbf{Alert Triage Performance Analysis:}
Although we couldn't evaluate \Sys's anomaly scoring on ATLASv2 due to the lack of necessary historical frequency information from the alert data, we assessed its clustering capability against NoDoze and RapSheet. All systems were trained using benign data from Day 1 to Day 4 of the ATLASv2 dataset and tested on Day 5, which comprises both benign and attack data. The performance, as illustrated in Figure~\ref{figauc}, shows CarbonFilter with an AUC of 0.94, significantly surpassing NoDoze (AUC of 0.60) and RapSheet (AUC of 0.90). This evaluation underscores CarbonFilter's scalability and its precision in alert triage, demonstrating its superior capability in distinguishing between false positives and true threats efficiently.

Additionally, we evaluated the timing overhead when incorporating provenance-based methods on top of \Sys. Initially, \Sys triaged all alerts, clustering them effectively; it suppressed 317 out of 491 alerts in ATLASv2. The remaining 174 alerts were then processed using provenance-based approaches. This process is detailed in the last two rows of Table~\ref{tab:computational-overhead}. Notably, the time per alert was similar, yet overall processing time significantly decreased. Specifically, employing NoDoze alone led to a 5,064 fold increase in total time overhead. In comparison, while integrating NoDoze with \Sys improved the performance by $\sim 3$ times, it still resulted in 1,798 fold increase in time overhead as compared to using \Sys by itself. This improvement, albeit still a substantial time overhead, highlights the efficiency of merging \Sys with provenance-based techniques to lower total alert processing time. 

\section{Related Work}
\label{sec:related_work}

In Section~\ref{sec:background}, we discussed the limitations of existing causality-based techniques concerning scalability. We continue our discussion of related work here.

\par \textbf{Causality-Based Alert Correlation} - Causality-based alert correlation techniques leverage information flow among various system entities to deduce alert correlations. The primary work in this area was pioneered by Zhai et al.~\cite{zhai2006integrating}. They were the first to use kernel-level audit logs to correlate different alerts. In their work, they first create dependency graphs from kernel-level audit logs, which are subsequently utilized for correlating threat alerts. HOLMES\cite{holmes} also makes use of data provenance and dependency graphs to establish correlations between different alerts. Their approach's central insight is that while the specific attack steps may greatly vary among different Advanced Persistent Threats (APTs), high-level APT behavior often aligns with the APT kill-chain.

\par \textbf{Causality-Based Alert Triage} Several causality-based alert prioritization systems have been presented in the literature~\cite{fang2022back, NODOZE, swift, hbm2020}. DEPIMPACT~\cite{fang2022back}, NoDoze ~\cite{NODOZE} and Swift~\cite{swift} all prioritize alerts based on their anomalous contextual history.  At a broader level, these systems initially assign anomaly scores to individual edges in the provenance graphs, considering different features such as prior event frequency, timing, data flow amount, and node degree, among others. Subsequently, they aggregate the anomaly scores along neighboring edges in the provenance graph. Ultimately, the final anomaly scores are used for alert prioritization. These systems often face challenges with stealthy attacks like living-off-the-land attacks, as attackers can potentially employ benign applications for malicious actions, thereby avoiding any logging of malicious low-level activities~\cite{inam2023sok}.  To triage such alerts by sophisticated attacks, RapSheet~\cite{hbm2020} proposes a threat scoring scheme that assesses each alert's severity based on causal dependencies between threat alerts, rather than encoding low-level dependencies.

While causality-based approaches have shown promise in small-scale evaluation test beds, they face scalability challenges that hinder their applicability to the demands of commercial EDR systems.



 \section{Discussion \& Future Work}
 \label{sec:future_work}

While \Sys offers an effective and scalable solution for alert triage, we foresee opportunities for further improvement. The system outlined here effectively triages alerts as seen today (including commodity attacks and living-off-the-land attacks). However, similar to prior systems, our approach might encounter challenges when dealing with highly sophisticated attacks, such as supply chain attacks (e.g., SolarWinds) which emulate nearly all aspects of normal behavior. However, because \Sys is highly granular compared to the majority of anomaly detection systems, we plan to leverage this granularity in the future.

Provenance-based approaches have displayed promise in the realm of alert triage and in the detection of more sophisticated attacks. However, their scalability remains a significant challenge.
We intend to explore computationally efficient methods for integrating provenance-based approaches into the existing framework of \Sys. A significant challenge in this regard is determining whether provenance-based techniques can be adapted to meet the scalability requirements of commercial EDRs.

We utilize TLSH similarity digests due to the design considerations discussed in Section~\ref{SecUsingTLSHforClustering}. We do not make use of certain embedding algorithms (e.g., doc2vec, etc.) for similarity computation, as these algorithms are less scalable during inference due to their reliance on vector calculations. Moreover, training the embedding model itself poses technical challenges. In contrast, while TLSH is computationally efficient, it has its limitations, including a minimum string size requirement of 50 bytes. In the future, we intend to explore more sophisticated embedding algorithms and adapt them to meet commercial scalability demands. Given the granularity of \Sys, it can easily be adapted to using an embedding algorithm instead of TLSH. 

The HAC-T clustering algorithm assigns a single centroid to each cluster. In certain instances, a single centroid may not adequately represent the entire cluster. For such cases, we plan to experiment with using multiple centroids for a single cluster in the future.

\section{Conclusion}
\label{sec:conclusion_future_work}

In this work, we address the significant challenge of managing alert fatigue in commercial EDR systems, focusing on the triage of tens of millions of alerts daily— a volume that surpasses the capabilities of existing alert triage methods documented in literature.

We introduce \Sys, a system that combines statistical and granular learning to accurately model endpoint activity behavior. \Sys distinguishes itself by offering a six-fold improvement in the Signal-to-Noise ratio without compromising on alert triage performance. Our approach stands out for its scalability, presenting a viable and effective solution for managing alert volumes for large organizations. 






\newpage


\bibliographystyle{plain}
\bibliography{references, bates-bib-master}

\newpage


\section{Supplemental Material}
\label{sec:supplement}

\subsection{Example of TLSH distance on command lines}
\label{subsec:tlsh_dist_example}

Table \ref{tab:TLSH-examples} shows three example command lines with TLSH digests T1, T2 and T3 respectively.\footnote{The actual TLSH hash is more than 70 characters long. For demonstration purposes, we replace the digests with labels T1, T2 and T3.}
The command line T1 is known to be a suspicious command based on feedback from threat analysts. On visual inspection, it can be noted that the command lines for T2 and T3 are very similar to each other. This observation is confirmed by comparing the distance between TLSH digests, where the distance between command lines (23) is an order of magnitude less than their respective distances from T1 (371 and 380).

\begin{table*}[htb]
\centering
\small
\caption{Examples of command lines and the TLSH distances between those command lines. \\}
\label{tab:TLSH-examples}

\begin{tabular}{|m{11.5cm}|c|}
\hline
\textbf{Command} & \textbf{TLSH} \\
\textbf{line} & \textbf{Label}
\\
\hline
\begin{verbatim}
"C:\Windows\System32\rundll32.exe" shwebsvc.dll,AddNetPlaceRunDll
\end{verbatim}  & T1
\\ \hline

\begin{verbatim}
"C:\WINDOWS\System32\WindowsPowerShell\v1.0\powershell.exe"
((New-Object System.Net.WebClient
).OpenRead(`https:\\www.google.com')).CanRead
\end{verbatim}  & T2
\\ \hline

\begin{verbatim}
"C:\WINDOWS\System32\WindowsPowerShell\v1.0\powershell.exe"
((New-Object System.Net.WebClient)
.OpenRead(`https:\\www.microsoft.com')).CanRead
\end{verbatim}  & T3
\\ \hline
\end{tabular}

\vspace*{5mm}

\begin{tabular}{|c|c|c|c|}
\hline
\textbf{TLSH Label} & \textbf{Dist T1} & \textbf{Dist T2} & \textbf{Dist T3} \\
\hline
T1 & 0 & 371 & 380 \\ \hline
T2 & 371 & 0 & 23 \\ \hline
T3 & 380 & 23 & 0 \\ \hline
\end{tabular}
\end{table*}

\newpage

\begin{figure*}[thb]
    \centering
    \includegraphics[width=5.0in]{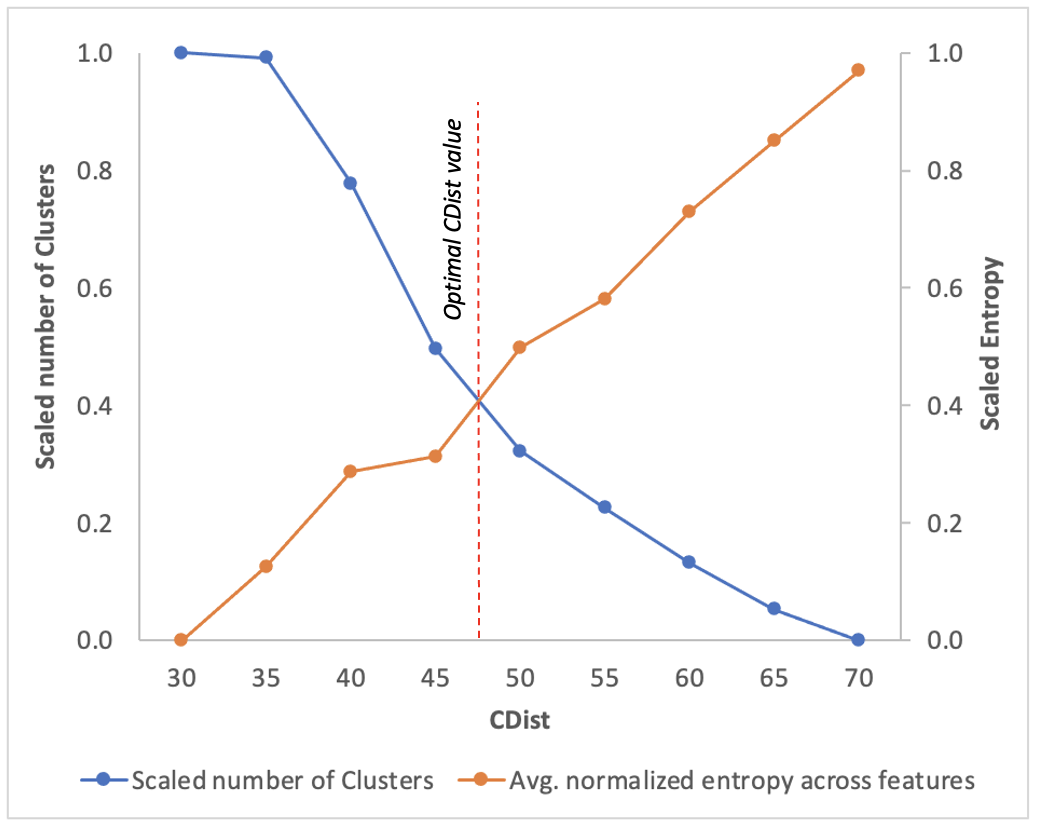}
    \caption{Relationship of $CDist$ with the number of clusters and average entropy across behavioral features.}
    \label{fig:trade-off-chart}
\end{figure*}

\subsection{Choosing Optimal CDist}
\label{subsec:ChoosingoptimalCDist}

The optimal value for $CDist$ is a trade-off between the number of clusters and the homogeneity within each cluster. To obtain the optimal value of $CDist$ for our alert data, we test various $CDist$ values in the range 30 -- 70 based on best practices established in \cite{oliver2013TLSH}. 

\begin{enumerate}
    \item \textbf{Number of Clusters:} During the real-time inference, a search across all the cluster centroids is performed using ANN ($\S$ \ref{subsec:evaluation_ann}). The time complexity of this search is of the order of $O(log(n))$, where n is the number of cluster centroids. Therefore, we have a preference for a smaller number of clusters. We run separate clustering jobs for all the command lines using $CDist$ value 30 --70 (in increments of 5) and compute the number of clusters for each value of $CDist$. With an increase in the value of $CDist$, the number of clusters decrease as larger clusters are being formed. 
    
    \item \textbf{Homogeneity within each cluster:} As the value of $CDist$ increases, larger clusters get formed, which leads to a reduction in homogeneity within the clusters. We use Shannon's entropy \cite{6773024} to measure the homogeneity for the clusters; the smaller the value of entropy, the higher the homogeneity.
    We measure the entropy change across various security features of both the actor process and the parent process, including process reputation, privilege escalation, digital signature states, process path and process user name.
    With an increase in the value of $CDist$, the entropy increases as larger clusters are being formed. 
    
\end{enumerate}

To measure the trade-off between the number of clusters and entropy, we randomly sampled 5 million alerts from the alerts population.
For each value of $CDist$, we have a set of cluster centroids.
For each value of $CDist$, we assign the 5 million test alerts to the closest centroid and compute the entropy for the security features defined above. We scale both the entropy values and the number of clusters into the range $[0, 1]$ using min-max normalization \cite{DBLP:journals/corr/PatroS15}.
Fig.~\ref{fig:trade-off-chart} shows how the increase in $CDist$ value impacts these two metrics. It was determined that a $CDist$ value between $45$ to $50$ would help us to maximize cluster size while ensuring cluster homogeneity on our data. For the purpose of further evaluations shown in this paper, we chose $ CDist ~=~ 50$.



\end{document}